\documentstyle[a4wide,epsf,11pt,titlepage]{article}

\newcounter{nref}
\newcommand{\bbib}{%
  \renewcommand{\refname}{\large\bf References}%
  \begin{thebibliography}{99}%
  \setcounter{enumiv}{\thenref}}
\newcommand{\ebib}{%
  \end{thebibliography}%
  \setcounter{nref}{\arabic{enumiv}}}
\newcommand{\head}[3]{%
  \setcounter{nref}{0}%
  \thispagestyle{empty}%
  \section*{\LARGE\bf #1}%
  \stepcounter{section}%
  \addcontentsline{toc}{section}{#1}%
  \large\itshape%
  #2\\\vspace{0.1pt}\\%
  #3%
  \normalsize\upshape%
  \bigskip}

\begin{document}

\let\jnlstyle=\rm
\def\refjnl#1{{\jnlstyle#1\/}}
\def\aj{\refjnl{Astron.J.}}
\def\apj{\refjnl{ApJ}}
\def\apjl{\refjnl{ApJ.Lett.}}
\def\apss{\refjnl{Ap.Space~Sci.}}
\def\aaps{\refjnl{Astron.Ap.Suppl.}}

\def\e#1{$\times 10^{#1}$ }
\def\ee#1{$10^{#1}$ }
\def\msun{M_\odot}
\def\Msun{$\msun$}
\def\rsun{R_\odot}
\def\Rsun{$R_\odot$}


\head{Shock breakouts in SNe~Ib/c}
     {S.I.\ Blinnikov$^1$, D.K.\ Nadyozhin$^1$,
     S.E.Woosley$^{2}$ and E.I.Sorokina$^{3}$}
     { $^1$ Institute for Theoretical and Experimental Physics,
            Moscow\\
$^2$ UCO/Lick Observatory,  University of California, Santa Cruz\\
$^3$ Sternberg Astronomical Institute, Moscow
      }



It was realized long ago that a shock propagating
along the profile of decreasing density should accelerate
\cite{blinn.GFK56,blinn.Sak60}.
In compact presupernovae, like SNe~Ib/c,
the shock can become relativistic \cite{blinn.JM71} and
is able to produce a burst of X-ray and even $\gamma$-ray
radiation  \cite{blinn.Col69,blinn.BKINC75,blinn.WooAA93}. The case
of a peculiar type Ic SN1998bw, probably related to GRB980425,
has shown that SNe~Ib/c may have the highest explosion energy
and highest production of $^{56}$Ni among core-collapsing supernovae
\cite{blinn.Iwamoto,blinn.WooES99}. 
Supernovae of type Ib/c are also interesting for theory due to
problems with their light curve modeling. Their understanding
may serve as a diagnostics of mass loss from massive stars at
the latest phases of evolution.

Numerical modeling of shock breakout in SNe Ib/c
was done previously using some simplifying
approximations \cite{blinn.ensmanphd}. Our  method, realized
in  the code  {\sc stella}, allows us to get more reliable
predictions for the outburst.
Improvements in the theory done in
the current work:
\begin{itemize}
\item
multi-energy-group time-dependent radiation transfer,
\item
taking into account of (some) relativistic effects.
\end{itemize}

\begin{figure}[ht]
   \centerline{\epsfxsize=0.6\textwidth\epsffile{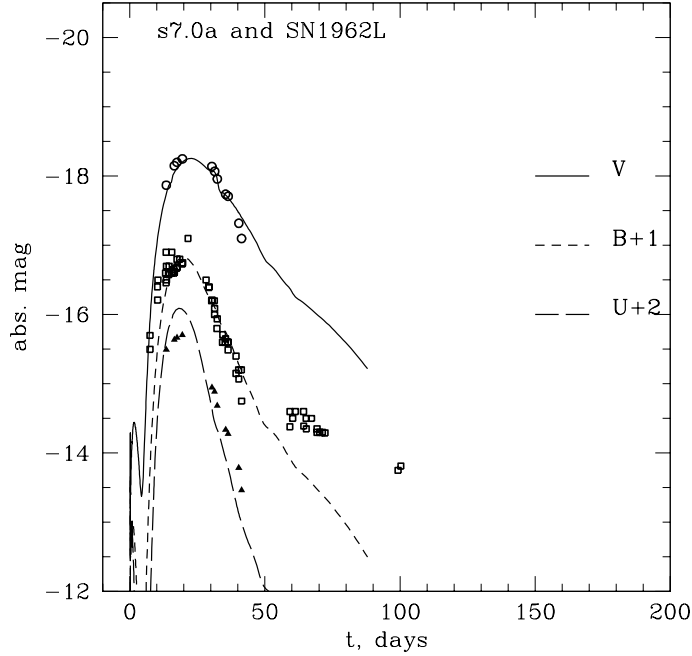}}
  \caption{Theoretical UBV fluxes for the model 7A, exploded
  with $E_{\rm kin}=1.47\times 10^{51}$ ergs in comparison
    with observations of SN1962L of type Ib.}
  \label{blinn.mags62l}
\end{figure}

A representative presupernova in our runs was a WR star built by the
code {\sc kepler} \cite{blinn.woolangw2} (model 7A).
Late light curves and spectra for this
model were studied in  \cite{blinn.Wee97}.
We present UBV light curves found for the same model
by {\sc stella} in Fig.~\ref{blinn.mags62l}.

The Model 7A has mass 3.199 \Msun (including the mass of the
collapsed core). Its radius prior to explosion is not strictly
fixed because the outer mesh zones in {\sc kepler} output
actually model the strong stellar wind and are not
in hydrostatic equilibrium.
So,  we fixed the radius by hand and
we got four models with radii from 0.76 up to 2 \Rsun\ which  {\em are}
in hydrostatic equilibrium. Our results are summarized in the Table.

\begin{figure}[ht]
  \centerline{\epsfxsize=0.6\textwidth\epsffile{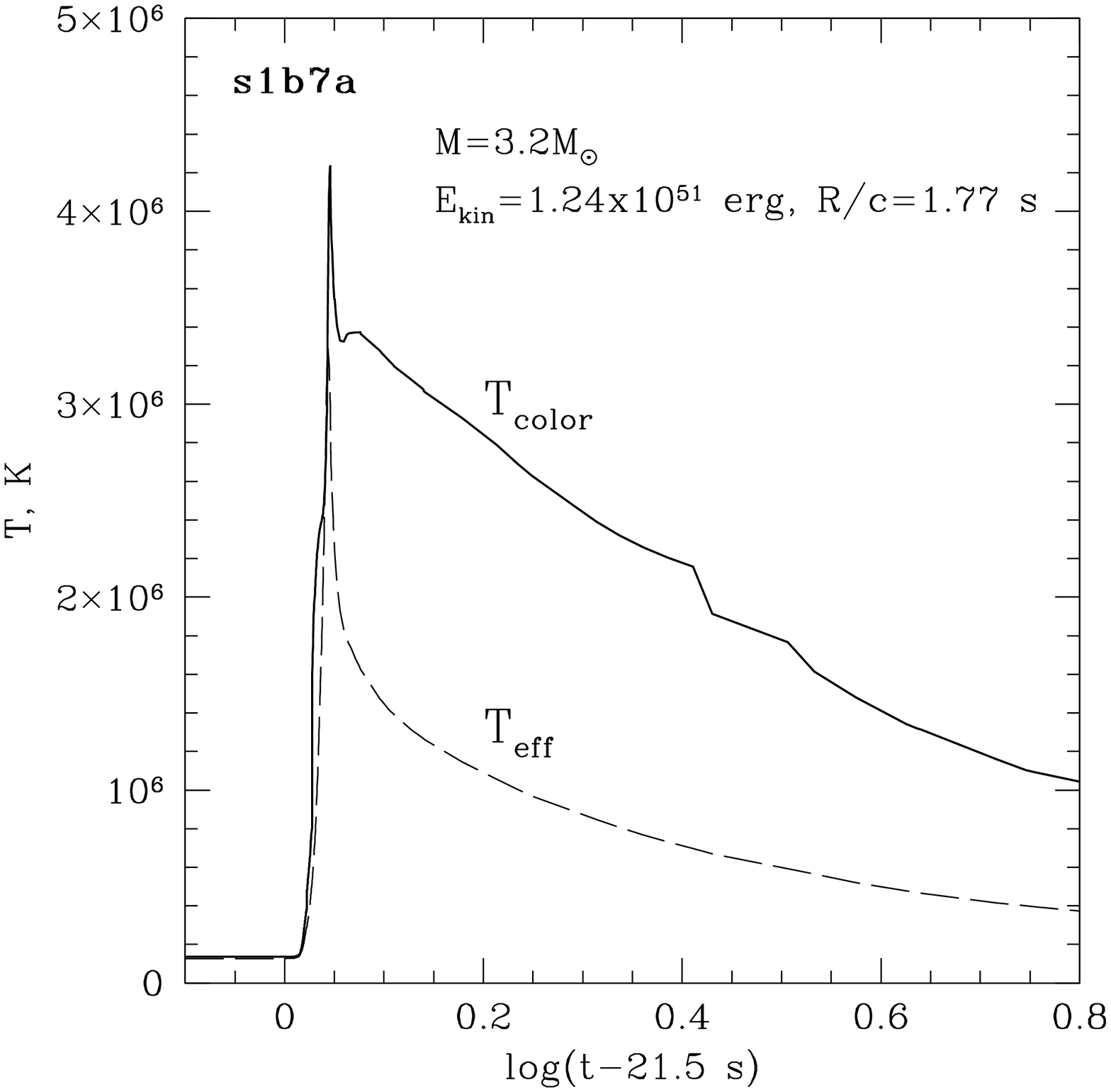}}
  \caption{Effective and color temperatures of emerging
  radiation in one of the runs.}
  \label{blinn.lTbTe}
\end{figure}

Explosions with {\sc kepler} \cite{blinn.Wee97}
gave maximum temperature of photons $T \sim 5\times 10^5$ K
at shock breakout. We have much finer mesh zoning at the edge of
the star (down to 1\e{-12} \Msun) and better physics and we find
higher values of $T$. The Fig.~\ref{blinn.lTbTe} shows
the difference between effective and color temperatures (labels `e'
and `c', respectively, in the Table).
One should note that the peak values of
luminosity and temperature given in the table and plots
in Fig.~\ref{blinn.lTbTe} do not contain the
light travel time correction.
\begin{figure}[ht]
 \centerline{\epsfxsize=0.45\textwidth\epsffile{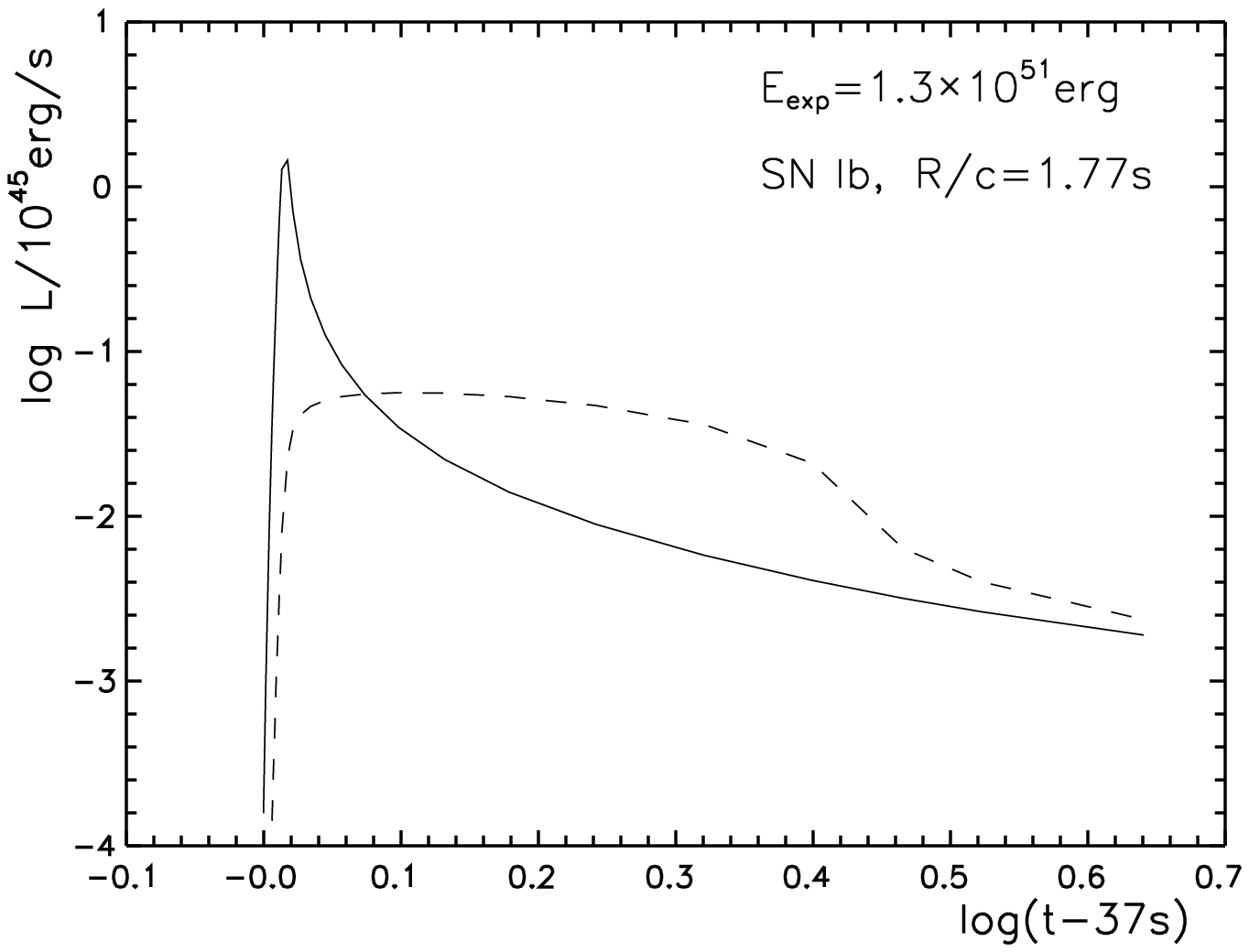}
  \epsfxsize=0.45\textwidth\epsffile{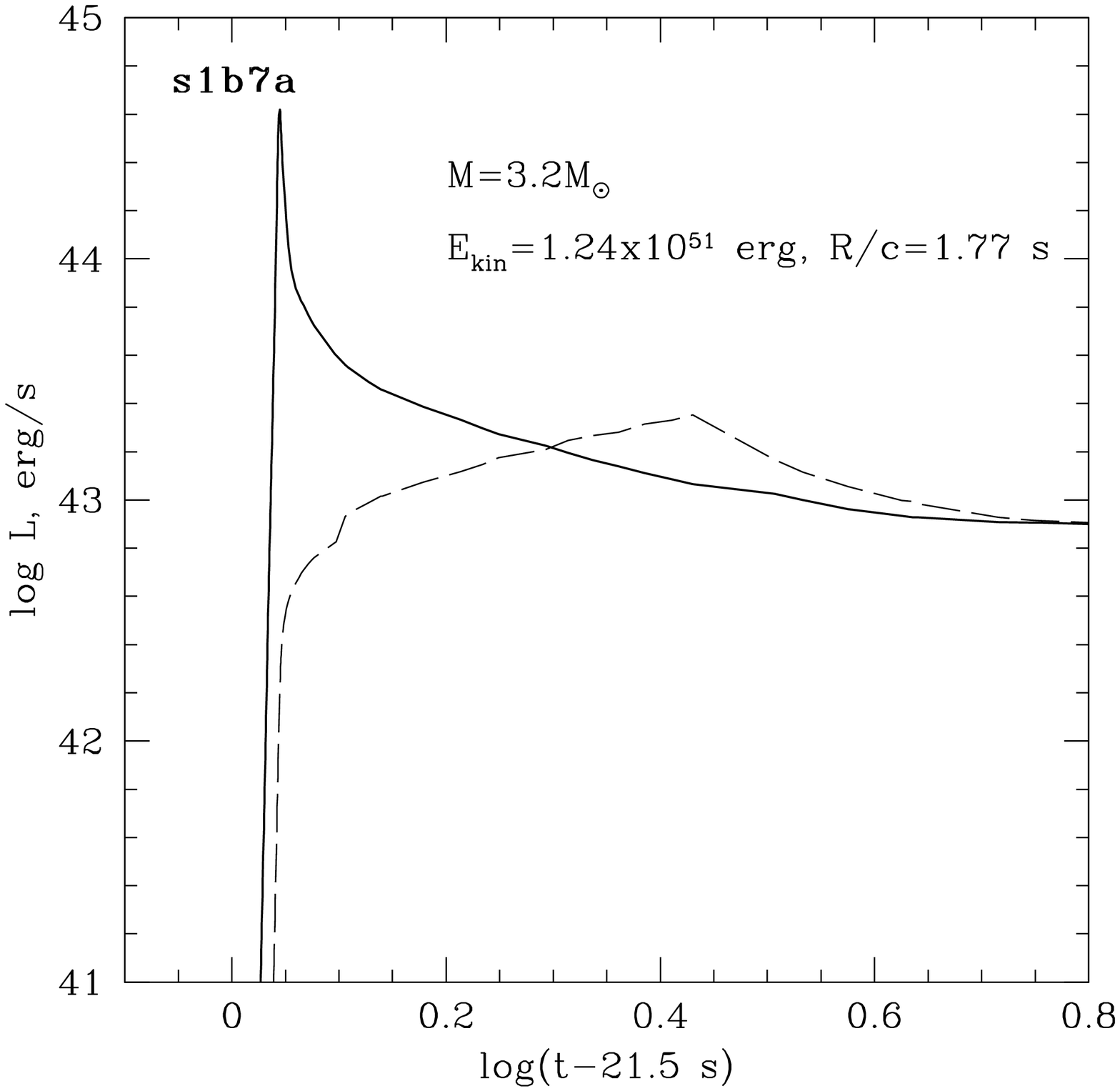}}
  \caption{The left figure shows the results for shock breakout
  luminosity  by the equilibrium radiation diffusion hydrocode {\sc snv};
  the right one: a similar presupernova model, results obtained by
  {\sc stella}. Dashed lines show the light travel time corrected
  luminosity. In case of {\sc stella} the correction is approximate.}
  \label{blinn.llumavg}
\end{figure}
Models, labeled as 3.5N were computed by D.K.Nadyozhin in 1992
with the equilibrium radiation diffusion hydrocode {\sc snv}.
Left plot in Fig.~\ref{blinn.llumavg} shows one of his results.
Dashed lines in Fig.~\ref{blinn.llumavg} demonstrate the effect
of averaging the light curve due the light travel time correction.

\begin{center}
\begin{tabular}{llllll}
\multicolumn{6}{c}{\large\bf Parameters of shock breakouts}\\
\hline \hline
 \\[-3mm]
 $M$ & $R_0$ & $E_{\rm kin}$ & $L_{\rm p}$   & $T_{\rm p}$ & $\Delta t$  \\
 \Msun & \Rsun & foe & erg/s   & $10^{6}$K & s  \\
\hline
3.2  &  0.76 &  1.24   & 4.2\e{44}  & 4.2c  &  0.021  \\ 
3.2  &  1.00 &  1.32   & 5.8\e{44}  & 4.3c  &  0.026  \\ 
3.2  &  1.23 &  1.30   & 6.8\e{44}  & 4.3c  &  0.043  \\ 
3.2  &  2.   &  1.39   & 9.4\e{44}  & 4.3c  &  0.12   \\ 
3.2  &  2.   &  4.36   & 3.6\e{45}  & 5.3c  &  0.028  \\ 
3.2  &  2.   &  8.86   & 4.8\e{45}  & 7.2c  &  0.020  \\ 
\hline
3.5N  &  0.76 &  1.30   & 1.4\e{45}  & 5.1e  &  0.028 \\ 
3.5N  &  1.23 &  1.30   & 8.1\e{44}  & 3.5e  &  0.067 \\ 
\hline
\end{tabular}
\end{center}
\noindent {\small
$M$ and $R_0$ --- presupernova mass and radius, respectively,
           in solar units\\
          $E_{\rm kin}$ --- the kinetic energy at infinity in $10^{51}$ ergs\\
          $L_p$ and $T_p$ --- the peak luminosity and temperature\\
           $\Delta t$ is the width of the
            light curves at 1 stellar magnitude below $L_p$ }


\bigskip

We conclude that photon spectra at shock breakouts in SNe~Ib/c
peak near $3kT \sim 1$ keV and the X-ray fluences predicted can be
observed by future missions up to tens of Mpc.

We are grateful to MPA staff for permanent and generous support.
Our work in US was supported by grants NSF AST-97 31569 and
NASA - NAG5-8128, in Russia RBRF 00-02-17230 and RBRF 02-02-16500.



\end{document}